# ON THE STRUCTURE OF THE FREE SURFACE OF SELF-BINDING QUARK MATTER

G. S. Hajyan   Yerevan State University, Armenia.
ghajyan@ysu.am

*The distributions of electrons and of electrical potential at the free surface of strange quark matter are determined within the framework of the MIT bag model. It is shown that, with allowance for the $\beta$ decay of quarks near the surface due to the outward escape of electrons, the electric charge density of quarks at the surface increases by a factor of 17-25, the thickness of the transitional layer decreases from 230 Fm to 15 Fm, and the field strength increases by a factor of 1.7. The difference between the chemical potentials of electrons at the surface and in deep layers decreases from 7 MeV to 0.8 MeV, which increases the limiting possible density of ordinary matter above a strange quark star.*

## 1. Introduction

The development of the theory of nuclear matter led to the hypothesis of the existence of a pion condensate [1], and later to that of strange quark matter (SQM) [2], which consists of almost equal numbers of $u$, $d$, and $s$ quarks, the electric charge of which is neutralized by electrons or positrons [3]. Depending on the numerical values of the physical constants of the theory, which are rather coarsely determined at present, SQM can be more bound than the matter of ordinary atomic nuclei.

At the free surface of such matter the density drops sharply from superdense values to zero, and due to the partial escape outward of electrons (positrons), a thin charged layer is formed, the field of which, in the case of electron neutralization of the quark electric charge, prevents the penetration of ordinary atomic nuclei into the SQM. This makes possible the virtually infinitely long coexistence of ordinary stellar matter and SQM in direct contact [4].

The electron distribution and the structure of the electrostatic field at the surface of superdense, self-confining nuclear matter has been studied in [5, 6] for pion nuclei and in [4, 7] for SQM. In those papers it was assumed that after the partial escape outward of electrons near the surface, the hadron component is not reorganized, and the electron distribution and electrostatic field potential were found by the Thomas–Fermi method. Under this assumption, the thicknesses of the charged quark layer and the outer electron cloud turn out to be on the order of 300 Fm and 1000 Fm, respectively.

The study of the properties of SQM is important for the physics of superdense celestial bodies. Quark matter may be formed and manifested in just such stars. In particular, the existence of a new branch of superdense objects is possible: strange stars, which differ fundamentally from ordinary neutron stars [4, 8, 9]. It was shown in [10] that the surface of a hot, bare strange star emits electron–positron pairs.

In the present work we determine the electron distribution and electric field potential at the free surface of self-confining, strange quark matter with allowance for its reorganization due to the β decay of $d$ and $s$ quarks.

## 2. Basic Equations

Both for strange quark nuclei and especially for a strange quark star, the thickness $l$ of the charged surface layer is very small relative to the curvature of the surface. We therefore consider the one-dimensional case. For ordinary nuclei the baryon density is almost constant over the entire nucleus and only in a surface layer ~2.5 Fm thick does it drop continuously to zero. We then assume that the analogous thickness $l_q$ for the quark case is no greater than this value. If we have $l \gg l_q$, then the change in quark density at the surface can be considered to be discontinuous in a first approximation. On the other hand, the increase in pressure and density in a surface layer with a thickness of several thousand fermis due to gravitational attraction is so small that this interaction can be neglected and the pressure can be assumed to equal zero in these "deep" layers.

In the Cartesian coordinate system let the half-space $x \leq 0$ be filled with self-confining, strange quark matter. In sufficiently deep layers ($|x| \gg l$) the matter is electrically neutral and has a zero pressure. Only at the very surface, in the region $x \leq l$, do some electrons escape into the space $x > 0$, forming an electron cloud.

We introduce the following notation: $e$ is the absolute value of the charge of an electron; $n_i$, $e \cdot q_i$, and $\mu_i$ are the concentration, electric charge, and chemical potential of the $i$-th type of particle, respectively; $V$ is the electrical potential multiplied by $e$ (henceforth the potential); $V_{surf}$ is the value of $V$ at the surface (at the point $x = 0$).

All the parameters at zero pressure ($x = -\infty$) are designated by an additional zero subscript, while those at the surface ($x = 0$) are designated by the subscript surf.

The potential $V$ is determined by the Poisson equation

$$V'' = -4\pi e^2 (q_u n_u + q_d n_d + q_s n_s + q_e n_e),$$
$$q_e = -1, \quad q_u = 2/3, \quad q_d = q_s = -1/3, \tag{1}$$

with the boundary conditions

$$V'|_{x=-\infty} = V'|_{x=+\infty} = 0 \tag{2}$$

(there is no field at infinity, since the system as a whole is electrically neutral).

We also take a zero potential at infinity,

$$V|_{x=+\infty} = 0, \tag{3}$$

thus identifying $V(x)$ with the depth of the well of the electrostatic field for electrons at each point $x$.

The conditions under β equilibrium have the form

$$\mu_d = \mu_u + \mu_e, \tag{4}$$

$$\mu_d = \mu_s. \tag{5}$$

Adding to (4) and (5) the conditions of electrical neutrality and zero pressure, we determine $n_{0i}$ and $\mu_{0i}$. The electrochemical potential $\mu_i + q_i V$, which determines the chemical potential of particles at any point in terms of the potentials $V(x)$ and $V_0$ and in terms of the already known $\mu_{0i}$, remains constant over the entire system for each particle. We thus have

$$\left.\begin{array}{l}\mu_u(x)=\mu_{u0}+\dfrac{2}{3}[V_0-V(x)],\\[4pt] \mu_s(x)\equiv\mu_d=\mu_{d0}-\dfrac{1}{3}[V_0-V(x)],\\[4pt] \mu_e(x)=\mu_{e0}-[V_0-V(x)].\end{array}\right\} \qquad (6)$$

Following the Thomas–Fermi method, we assume that at each point the electron levels are filled to the value $V(x)$. We then obtain

$$\mu_e(x)=V(x)+1, \qquad (7)$$

i.e., in deep layers with $\mu_e \gg 1$ we have $V(x) = V_0 = \mu_{0e}$.

The concentrations are uniquely determined in terms of the chemical potentials:

$$n_i = n_i(\mu_i). \qquad (8)$$

In the region $x > 0$, where there are only electrons, Eq. (1) takes the simple form [1]

$$V'' = \frac{4e^2}{3\pi}\left(V^2+2V\right)^{3/2}. \qquad (9)$$

Here and below we use a system of units in which the electron's mass $m_e$, the speed of light $c$, and Planck's constant $\hbar$ are equal to unity.

Thus, Eqs. (1), (2), (3), (6), and (8) uniquely determine the function $V(x)$.

## 3. Solution of the Problem

The problem was solved within the framework of the SQM bag model, in accordance with which the masses of $u$ and $d$ quarks are neglected and the quark–gluon interaction is taken into account to first order with respect to the interaction constant $\alpha_c$.

In accordance with this model, the thermodynamic potentials $\Omega_i$, the pressure $P$, and the energy density $\varepsilon$ are determined from the equations [3]

$$n_i = -\frac{d\Omega_i}{d\mu_i} = -\mu_i^3(1-2\alpha_c/\pi)/\pi^2, \quad i=u,d,$$

$$n_s = -\frac{d\Omega_s}{d\mu_s} = \frac{\mu_s^2-m_s^2}{\pi^2}\times$$

$$\times\left(\left(\mu_s^2-m_s^2\right)^{1/2}-\frac{2\alpha_c}{\pi}\left(\mu_s-\frac{3m_s^2}{\left(\mu_s^2-m_s^2\right)^{1/2}}\ln\left(\frac{\mu_s+\left(\mu_s^2-m_s^2\right)^{1/2}}{\rho}\right)\right)\right)$$

$$n_e=\mu_e^3/3\pi^2, \quad P=-\sum_i \Omega_i - B, \quad \varepsilon = \sum_i(\Omega_i+\mu_i n_i)+B,$$

where $m_s$ is the mass of the $s$ quark, $B$ is the bag parameter, and $\rho = 313$ MeV is the renormalization constant for the mass of the $s$ quark. The circumstance cited in the introduction is related just to the uncertainty in the numerical values of these quantities.

For determinacy, we consider the numerical values of these constants for the case in which electrical neutrality is provided by electrons. Since they are ultrarelativistic ($V \gg 1$), the solution of (9) has the simple form

$$V(x) = V_{surf}/\left(1 + \sqrt{2/3\pi}\, eV_{surf} x\right). \tag{10}$$

The conditions of continuity of the potential and field strength together with (10) connect the derivative $V'$ at the surface with the value of $V_{surf}$:

$$V'_{surf} = -\sqrt{2/3\pi}\, eV^2_{surf}, \tag{11}$$

which is the boundary condition for (1) at the point $x = 0$.

Equation (1) is solved numerically by an eighth-order Runge–Kutta method. Integration starts from the point $x = 0$ with the value of $V_{surf}$ for which $V$ monotonically approaches $V_0 = \mu_{e0}$ for sufficiently large $x$ (thousands of fermis).

## 4. Results

The results of the calculations are given in Table 1 and in Fig. 1 (version b). For comparison, we also give the corresponding numerical values and dependences of the physical parameters for the model without allowance for quark reorganization (version a).

In Fig. 1 for the set $\alpha_c = 0.15$, $m_s = 200$ MeV, and $B = 60$ MeV·Fm$^{-3}$ we give the coordinate dependences of the potential and the density of uncompensated positive electric charge $(\rho^+ - \rho^-)/\rho^+_{surf}$ in the region of $-50\,\text{Fm} \leq x \leq 50\,\text{Fm}$ ($\rho^+$ and $\rho^-$ are the densities of positive and negative charges, respectively). As we see, the β decay of quarks, as expected, does not qualitatively change the picture, but there are significant quantitative differences: the thickness of the transitional

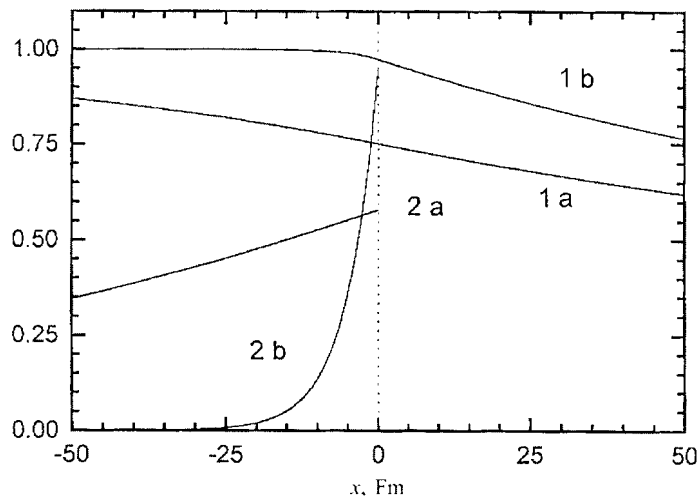

Fig. 1. Coordinate dependences of the potential $V/V_0$ (curves 1a and 1b) and uncompensated electric charge $(\rho^+ - \rho^-)/\rho^+_{surf}$ (curves 2a and 2b). Versions a and b are without and with allowance for quark reorganization in the outward escape of electrons, respectively. The quark matter fills the region $x \leq 0$.

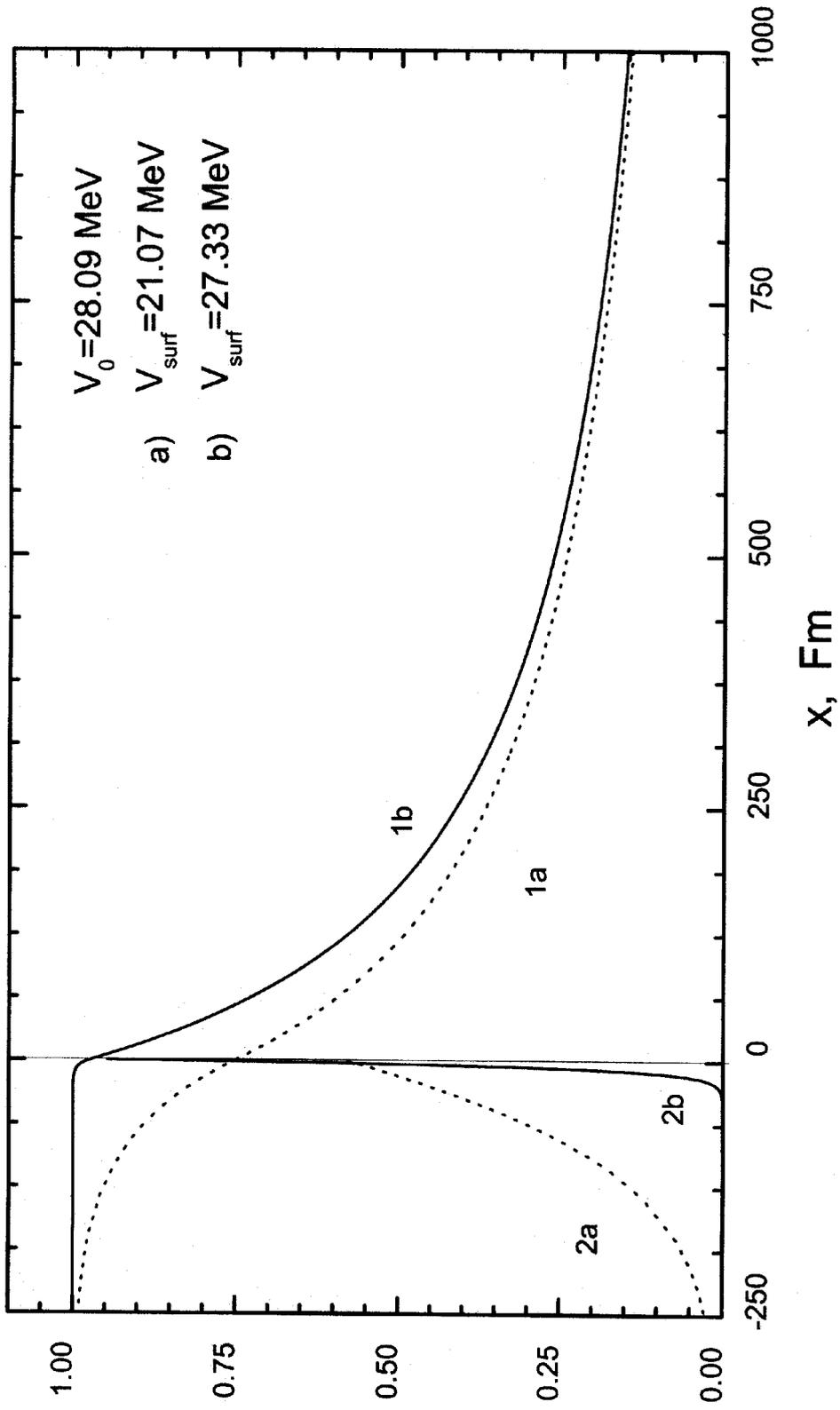

Fig.1b. Same dependences from fig.1 for coordinate interval -250Fm<x<1000Fm.

TABLE 1. Numerical Values of Some Physical Parameters of the Charged Layer Near the Surface of Strange Quark Matter*

| | $V_0$ MeV | $V_0$-$V_{surf}$ MeV | $E_{surf}$ V/cm | $\rho^+_{surf}/\rho^+_0$ | $\dfrac{\rho^+_{surf}-\rho^-_{surf}}{\rho^+_{surf}}$ | $\dfrac{n_0-n_{surf}}{n_0}$ | $l$ Fm | $L$ Fm |
|---|---|---|---|---|---|---|---|---|
| | | | $\alpha_c = 0.15$, $m_s = 200$ MeV, $B = 60$ MeV/Fm$^3$ | | | | | |
| a | 28.09 | 7.02 | 8.9 · 10$^{17}$ | 1 | 0.58 | 0 | 230 | 1100 |
| b | 28.09 | 0.76 | 1.5 · 10$^{18}$ | 17 | 0.95 | 1.8 · 10$^{-4}$ | 15 | 820 |
| | | | $\alpha_c = 0.05$, $m_s = 150$ MeV, $B = 50$ MeV/Fm$^3$ | | | | | |
| a | 18.87 | 4.72 | 4.0 · 10$^{17}$ | 1 | 0.58 | 0 | 340 | 1600 |
| b | 18.87 | 0.37 | 6.8 · 10$^{17}$ | 25 | 0.96 | 6.1 · 10$^{-5}$ | 16 | 1200 |

*Versions a and b are the results of calculations without and with allowance for quark reorganization in the outward escape of electrons, respectively, $E$ is the electric field strength, $\rho^+$ and $\rho^-$ are the densities of positive and negative electric charges, and $n = (n_u + n_d + n_s)/3$ is the baryon charge density.

quark layer is sharply reduced and the density of positive charge increases strongly.

The sizes of the regions in which 95% of the uncompensated positive and negative electric charges are concentrated are taken as the numerical values of the thickness $l$ of the transitional quark layer and the thickness $L$ of the outer electron cloud. As seen from Table 1, the thickness $l$ of the charged quark layer decreases by an order of magnitude (~15 Fm) due to β processes. The condition $l \gg l_q$ is still satisfied, however, and the neglect of the thin transitional quark layer at the very surface is justified.

If the reorganization of quarks at the surface is ignored, then we have $V_{surf} = 3V_0/4$ [4]. However, the β decay of $d$ and $s$ quarks fills the medium with "new" electrons in place of those that have escaped, thereby preventing a decrease in the electron chemical potential (see Fig. 1, curve 1b). This result is in complete accord with the Le Chatelier–Braun principle. In fact, the electron density in the SQM hardly changes. Here the density $\rho^+_{surf}$ of quark electric charge at the surface increases by 17 and 25 times, respectively, for the sets of values of $\alpha_c$, $m_s$, and $B$ considered.

In the quark transitional layer the SQM density becomes 0.01-0.02% lower than the density in deep layers.

It is easy see that the total charge $Q$ of external electrons above a unit area is $-V^2_{surf}/2\pi\sqrt{6\pi}$ while the thickness $L$ of the cloud is expressed in terms of the potential at the SQM surface: $L = \sqrt{30\pi}/eV_{surf}$. Despite the fact that the number of escaped electrons increases in the β decay of quarks, because the electron density increases everywhere, the thickness of this conventional layer decreases. This result, strange at first glance, is easily explained. With β decay the absolute charge increases on both sides of the surface, thereby increasing the field strength (see Table 1), which makes the system relatively more compact.

The presence of a strong electric field at the surface of a strange star makes possible the existence of two-layered, superdense celestial bodies [4] — strange stars, surrounded by an ordinary, degenerate, electron-nuclear plasma. The maximum density $\rho_{cr}$ of the crust can reach the value $\rho_{drip} \approx 4 \cdot 10^{11}$ g/cm$^3$ (the density of the onset of neutron drip from atomic nuclei). The free transfer of neutrons into the quark phase makes impossible the coexistence with SQM of ordinary degenerate stellar matter at a higher density than $\rho_{drip}$. This limiting value may be even lower, however, if we have $V_{surf} < \mu_e(\rho_{drip}) \approx 23$ MeV, the chemical potential of electrons in a degenerate electron-nuclear plasma at the density $\rho_{drip}$.

Then $\rho_{cr}$ is determined from the condition $\mu_e(\rho_{cr}) = V_{surf}$, when the charged dividing layer disappears altogether. Since allowance for quark reorganization increases the potential $V_{surf}$ by almost 25%, it is clear that the maximum density of the shell of ordinary degenerate matter also increases. The presence of such a shell on a low-mass strange star results in a strong increase in the star's radius (several hundred kilometers) [11], which facilitates their observation as intermediate objects on the Hertzsprung–Russell diagram between white dwarfs and neutron stars.

I wish to thank Prof. Yu. L. Vartanyan and participants in a seminar of the Department of the Theory of Wave Processes for a discussion of the results of this work.

This work was done under topic 2000-55, supported by the Ministry of Higher Education and Science of the Republic of Armenia.